\documentclass[]{copernicus}

\begin{document}
\nolinenumbers

\title{Experimental setup for synchronized surface and magnetic loss studies of grain oriented FeSi steel sheets}

\Author[1]{Korbinian}{Pfnür}
\Author[1]{Joachim}{Lüdke}
\Author[1]{Rainer}{Ketzler}
\Author[2]{Morris}{Lindner}
\Author[2]{Timmy}{Reimann}
\Author[2]{Benjamin}{Wenzel}
\Author[2]{Rocco}{Holzhey}
\Author[3]{Lev}{Dorosinsky}
\Author[1]{Franziska}{Weickert}

\affil[1]{Physikalisch-Technische Bundesanstalt,  
	Bundesallee 100, 38116 Braunschweig, Germany}
\affil[2]{Innovent e.V., Prüssigstraße 27b, 07745 Jena, Germany}
\affil[3]{TÜBITAK Gebze Yerleskesi, P.K. 54, 41470 Gebze/Kocaeli, Turkey}




\correspondence{Franziska Weickert (franziska.weickert@ptb.de)}

\runningtitle{TEXT}

\runningauthor{TEXT}

\received{}
\pubdiscuss{} 
\revised{}
\accepted{}
\published{}


\firstpage{1}

\maketitle

\begin{abstract}
We present technical details on an experimental setup that allows to measure magnetic losses in electrical steel sheets and the movement of magnetic domains on the sample surface simultaneously. The setup is suitable to investigate grain oriented electrical sheets in the polarization range 0.05\,T to 1.9\,T and at excitation frequencies between 50\,Hz and 4\,kHz. The screened surface area is 13\,mm x 18\,mm.   

\end{abstract}


\section{Introduction}

Power loss in soft magnetic materials has a negative impact on the economy of industrial societies. Since electrical steel sheets are used in generators, engines, and transformers of all kinds, even a small fraction of losses amounts to a huge total number, if all individual applications are considered together. Therefore from early on, a keen interest existed in the fundamental understanding of loss mechanisms in electrical steel sheets and a strong desire to optimize their properties under AC excitation.

To understand loss, it is necessary to study magnetic domain structures. Domains are formed in magnetic materials, because a local energy balance exists between exchange energy, magnetic anisotropy energy, magneto-elastic energy, and stray field energy [\cite{hubert_98}] that is non homogeneously distributed. Moreover, it creates domain patterns that have a more or less regular shape depending on the respective material. Static domain patterns themselves can be observed with a variety of different techniques [\cite{hubert_98}]. 

Magnetization and demagnetization processes of domains originating in domain wall motion and domain flips are a major causes of power loss in electrical steel sheets. Another mechanism is the induction of eddy currents. Depending on the specific type of steel, these contributions and dependencies on external parameters, such as excitation frequency, sample dimensions, and magnetic polarization vary. Early theoretical models, such as the one developed by [\cite{pry_58}] or the statistical model by [\cite{bertotti_98}] quantitatively describe experimental data with reasonable success. These models have been improved further, but a deeper understanding of domain dynamics must be gained by simultaneous studies of losses and domain movement.

Considering all observation techniques, only magneto optical (MO) methods that have been developed only recently, have recording times well below 1\,s. Low recording times allow in-situ investigations at frequencies at and above technical frequency of 50/60\,Hz. Another advantage of MO techniques are easy handling without elaborated sample preparation techniques as required, e.g. for Kerr microscopy. 

Here, we present an experimental setup that combines parallel measurement of magnetic loss with fast domain studies of grain oriented (GO) electrical steel sheets. 

\section{Technical realization}

\subsection{Loss measurements}

\begin{figure}[t] 
	\centering
	\includegraphics[width=9cm]{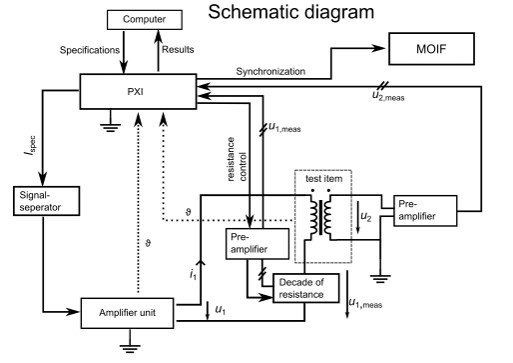}
	\caption{Schematics of the experimental setup for loss measurements $P$.}
	\label{fig1}
\end{figure}

At PTB, a high precision measurement setup exists for well defined polarization and frequency ranges between 0.05\,T - 2.3\,T, and 50\,Hz and 1\,kHz, respectively. It is optimized for measurements according to standard magnetic circuits, i.e. Epstein frames and single sheet testers (SST) [\cite{60404_2,60404_3}], leaving little flexibility in connecting magnetic circuits of non standard type. Therefore, it was necessary to develop a completely new experimental setup accommodating variable measurement frequencies from 50/60Hz to 5\,kHz. Moreover, the implementation of an amplifier with wide power, current, and voltage range was necessary to be able to connect non standard SSTs and other magnetic circuits. Later are required to have surface access to samples of various sizes, e.g. single Epstein strips of 30\,mm width with small total mass. A complex feedback control is used [\cite{luedke_01}] to achieve close to sinusodial signals and to obtain small measurement uncertainties and high reproducibility of the loss data. A schematic drawing of the setup is shown in figure\,\ref{fig1}. The measurement concept is based on an unloaded transformer with high precision voltage measurement across a dummy resistances $R_{1}$ on the primary and $R_{2}$ on the secondary side. Both values $U_{1}$ and $U_{2}$ are used to estimate the applied magnetic excitation field $H$ and the magnetic polarization $J$ of the material, and the power loss $P$. Further details of the measurement are described in [\cite{luedke_01}].

\subsection{Magnetic circuit}

\begin{figure}[h] 
	\centering
	\includegraphics[width=8cm]{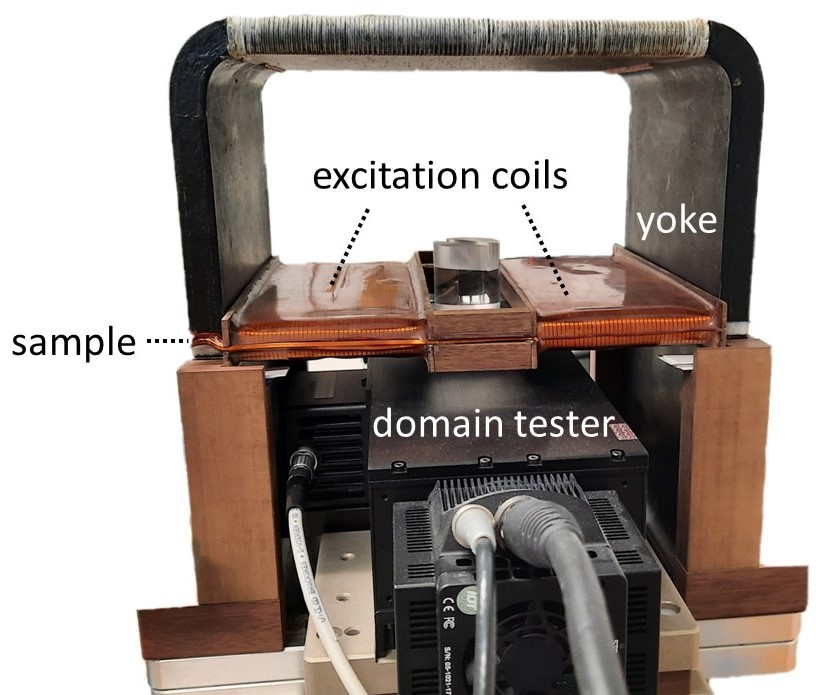}
	\caption{Magnetic circuit consisting of a yoke and a pair of excitation coils. The MO sensor of the domain tester is places in parallel under the sample (Epstein strip) between the coils.}
	\label{fig2}
\end{figure}

A picture of the magnetic circuit is shown in Fig.\,\ref{fig2}. It consists of a double yoke, a pair of excitation coils with pickup windings on top and the sample (electrical steel sheet) under investigation. The size of the double yoke is 30\,cm x 20\,cm. It consists of wrapped electric strips that are annealed to remove internal stress acquired during production. To reduce Eddy current induced losses directly in the yoke, the surface edges were cleaned by chemical etching after cutting the yoke into two halves. Sample dimensions are restricted by the yoke size. For this combination of yoke and sensor size (see below), Epstein stripes are most suitable for investigations. We manufactured 2 pairs of coils with different numbers of turns depending on the excitation frequency during the measurement. One pair with 100\,turns is best suitable for 50\,Hz to 200\,Hz measurements and the second pair consisted of 50\,turns each. Later one is  optimized for investigations at 300\,Hz to 5\,kHz range. 

The domain tester, black box in Fig.\,\ref{fig2} is placed within the yokes. We also implemented a hoist that maintains always a parallel surfaces between the domain tester sensor and the sample surface to avoid mechanical damage of the sensor during sample swap. 

\subsection{Domain tester}

\begin{table}[t]
\caption{Specifications of a Domain tester equipped with high speed 8\,kHz CCD camera and high power light source.} 
	\label{tab_domaintester}
	\begin{tabular}{lr}
		\tophline
		sensor type&A with in-plane sensitivity\\
		sensor area & 18\,mm x 13\,mm\\
		spacial resolution& 31$\mu$m\\
		frame rate & 8\,kHz\\
		camera resolution & 15.4\,$\mu$m\\
		minimum exposure time & 125\,ms\\
		power light source & 5\,W\\
		gray value depth & 10\,Bit\\
		camera interface & Gigabit Ethernet\\
		video output & MP4 format\\
		excitation frequency & up to 400\,Hz\\
		\bottomhline
	\end{tabular}
\end{table}

The domain tester is a recent development by Innovent e.V. It combines CMOS-MagView technology based on MO indicator films with a fast 8\,kHz camera in order to take images of moving domains. A high intensity LED light source is implemented to accommodate exposure times <\,200\,ms. A MO sensor type A resolving magnetic stray fields perpendicular to the sample surface is mounted inside the senor head. Its MO sensitive layer is made out of ferrimagnetic Bismuth-substituted iron garnet, which imposes a particularly strong Faraday rotation on passing light. The MO layer is deposited by liquid phase epitaxy on single crystalline substrate, which ensures gray scale imaging of high-resolution. The sensor size is about 13\,mm x 18\,mm and it is positioned close to the sample surface. Best imaging results are obtained for distances between sample and sensor of only a few $\mu$m, because magnetic stray fields decay rapidly with increasing distance. Further specifications of the domain tester are summarized in table\,\ref{tab_domaintester}.

In the combined setup, the AC excitation of the loss measurement is synchronize with the high-speed camera by a trigger signal from the PXI system as shown in Fig.\,\ref{fig1}. The trigger can be set in different configurations, e.g. at zero crossing of the excitation voltage.

\section{Results}

\begin{figure}[t] 
	\centering
	\includegraphics[width=8.3cm]{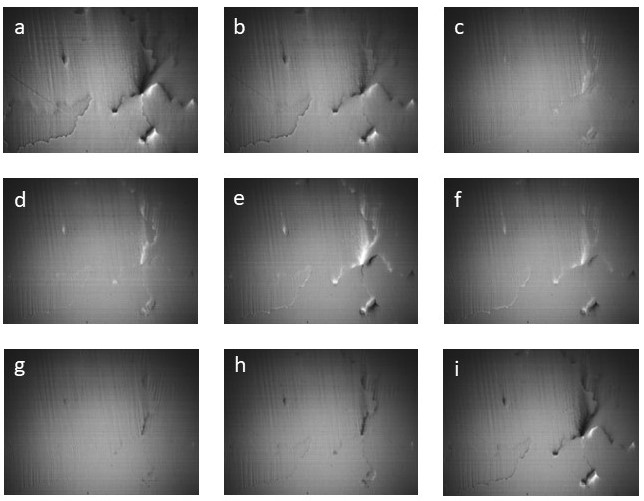}
	\caption{Sequence of images a) to i) taken by the domain tester at 1\,kHz excitation rate and 8\,kHz camera frame rate. Maximum and minimum polarization $J = \pm\,1.73$\,T of the Epstein strip is reached at image c) and g), respectively.}
	\label{fig3}
\end{figure}

The spacial resolution of the sensor is limited by the internal domain size of the MO film and the pixel size of the high speed CCD camera. Therefore, the setup is most suitable to investigate GO electrical steel sheets. Figure\,\ref{fig3} shows domain images taken for 1\,kHz sinusoidal excitation with maximum frame rate of 8\,kHz. i.e. sequences of 8\,images per one period are recorded. The exposure time per picture is set to 120\,$\mu$s. Maximum polarization of 1.73\,T is reached in c) leading to mostly aligned domains within the plane that do not exhibit strong stray field contributions. Every 8 images (compare images a) and i)), the domain pattern is repeated.

Successful tests have been carried out also at higher excitation frequencies of up to 4\,kHz. Here, the grey scale contrast is significantly reduced due to shorter integration times of 50\,$\mu$s and less.   

\section{Summary \& conclusion}
We build and successfully tested an experimental setup, where loss measurements and surface analysis on GO electrical steel sheets can be carried out simultaneously. We implemented a domain tester in a fully digital setup for loss measurements. Electrical steel sheets with sample sizes of up to 100\,mm width can be investigated in the frequency range up to 4\,kHz. With increasing detection frequency of the domain tester, image contrast is reduced due to shorter exposure times. 

Future work will include efforts for precise calibration of the loss setup. Furthermore, more advanced software tools need to be developed that allow averaging of images as used in stroposkopic surface analysis, e.g. Kerr microscopy by [\cite{hubert_98}]. This new capability will help to shed light on the relationship between domain dynamics and loss effects, especially at high frequencies in the kHz range that are not yet fully understood.










\authorcontribution{K.P. designed and build the loss measurement setup with contributions from J.L.. B.W., M.L., and R.H. designed and build the domain tester. K.P. combined both setups. K.P. and F.W. obtained experimental data. All authors discussed the results and provided contributions for the improvement of the setup. F.W. and L.D. drafted the manuscript with contributions from all authors.} 

\competinginterests{The authors declare that they have no conflict of interest.} 


\begin{acknowledgements}
This research work was partially supported by the 19ENG06 HEFMAG project, which was funded by the EMPIR program, and co-financed by the Participating States and the European Union’s Horizon 2020 research and innovation program.
\end{acknowledgements}






\bibliographystyle{copernicus}
\bibliography{all_202305.bib}

\end{document}